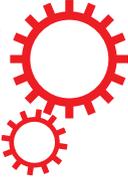

# Development of a computer-aided design software for dental splint in orthognathic surgery

Xiaojun Chen[1], Xing Li[1], Lu Xu[1], Yi Sun[2,3], Constantinus Politis[2,3] & Jan Egger[4,5]

In the orthognathic surgery, dental splints are important and necessary to help the surgeon reposition the maxilla or mandible. However, the traditional methods of manual design of dental splints are difficult and time-consuming. The research on computer-aided design software for dental splints is rarely reported. Our purpose is to develop a novel special software named EasySplint to design the dental splints conveniently and efficiently. The design can be divided into two steps, which are the generation of initial splint base and the Boolean operation between it and the maxilla-mandibular model. The initial splint base is formed by ruled surfaces reconstructed using the manually picked points. Then, a method to accomplish Boolean operation based on the distance filed of two meshes is proposed. The interference elimination can be conducted on the basis of marching cubes algorithm and Boolean operation. The accuracy of the dental splint can be guaranteed since the original mesh is utilized to form the result surface. Using EasySplint, the dental splints can be designed in about 10 minutes and saved as a stereo lithography (STL) file for 3D printing in clinical applications. Three phantom experiments were conducted and the efficiency of our method was demonstrated.

Orthognathic surgery procedures are mainly adopted to correct skeletal angle class II and III deformities, dent maxillofacial deformities, mandibular laterognathia, and maxillofacial asymmetries[1–4]. Computer-aided design has been gaining popularity in orthognathic surgery for enhancing the surgical treatment procedures and outcomes. Previous studies have demonstrated that computer-aided design and manufacturing (CAD/CAM) technology can improve the treatment[5–8]. For example, Lin *et al.*[9] proposed a method for producing customized positioning guides combined with the single-splint technique for orthognathic surgery. Zinser and Zöller[10] developed and patented a CAD/CAM technique for the fabrication of multiple surgical splints for orthognathic surgery. Schouman *et al.*[11] carried out a cadaveric study to evaluate the accuracy of the CAD/CAM generated splints in orthognathic surgery by comparing planned with post-operative 3D images. The study demonstrated that maxillary repositioning can be accurately approximated and predicted by specific computational planning and CAD/CAM generated splints in orthognathic surgery. Nimeskern presented a new web service based on software solution to obtain surgical splints for orthognathic surgery[12].

Obviously, the CAD softwares play a critical role in computer assisted orthognathic surgery. Nowadays, the softwares utilized in some studies for the generation of the splint are listed as below[13–15]:

(1) Simplant OMS 10.1 (Materialise®, Leuven, Belgium).
(2) Unigraphics NX 7.5 (Siemens PLM Software, TX, USA).
(3) Maxilim 2.2.2 (Medicim NV, Mechelen, Belgium).

However, the softwares mentioned above are commercial and the involved algorithms are not reported. Especially, as some of them are general CAD softwares, the process for splint generation is always time consuming and complicated. Therefore, the development of robust methods to support the design of the splints is of utmost importance.

[1]Institute of Biomedical Manufacturing and Life Quality Engineering, State Key Laboratory of Mechanical System and Vibration, School of Mechanical Engineering, Shanghai Jiao Tong University, Shanghai, China. [2]OMFS IMPATH Research Group, Department of Imaging and Pathology, Faculty of Medicine, KU Leuven, Leuven, Belgium. [3]Department of Oral and Maxillofacial Surgery, University Hospitals Leuven, Leuven, Belgium. [4]Faculty of Computer Science and Biomedical Engineering, Institute for Computer Graphics and Vision, Graz University of Technology, Graz, Austria. [5]BioTechMed-Graz, Graz, Austria. Correspondence and requests for materials should be addressed to X.C. (email: xiaojunchen@163.com) or J.E. (email: egger@icg.tugraz.at)

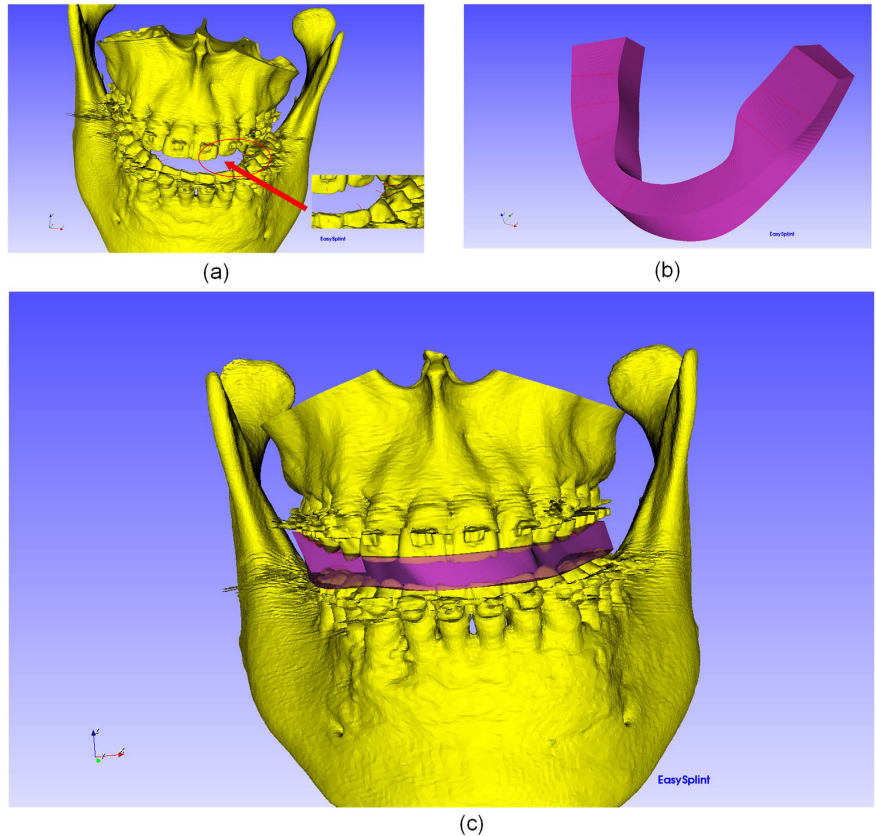

**Figure 1.** The generation of initial splint: (**a**) indicating the points to generate the lines. (**b**) The generated initial splint based on the lines. (**c**) The maxilla-mandibular model with the initial splint.

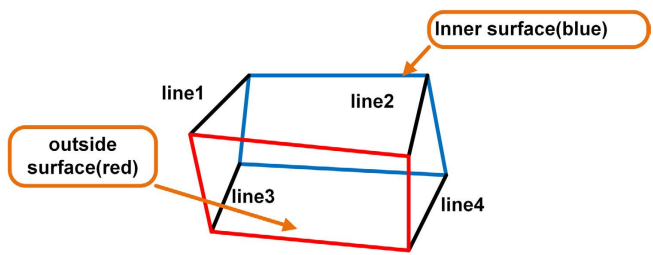

**Figure 2.** The reconstruction of the ruled surface, line1 and line2 are generated from maxilla model, line3 and line4 are generated from mandibular model.

In this study, a novel software named EasySplint for the design of the splint has been developed. Compared with the traditional softwares, an easier and more efficient approach to design the splint is adopted in EasySplint. The customized splint with specific shape can be achieved on the basis of the Boolean operation between the initial splint and maxilla-mandibular model, and the interference elimination.

## Methods
**Overview.** The architecture of the software is divided into the following four modules:

1. The module for preprocessing: The CT data of a patient can be imported into EasySplint to reconstruct the maxilla and mandibular models based on image segmentation, region growing and marching cubes algorithm. These two models can be repositioned according to clinical requirement.
2. The module for generation of the initial splint base: An array of points on the maxilla and mandibular model are manually selected. The positions and normals of these points are then utilized to generate lines, whose nodes then can be used to reconstruct ruled surfaces. The initial splint base is formed by appending the ruled surfaces. In addition, the positions and normals can be adjusted to better form the splint.
3. The module for Boolean operation: The type of Boolean operation here is the difference between initial splint model and maxilla-mandibular model. The algorithm of our Boolean operation is to split the intersected mesh, then to assemble related mesh to form the final dental splint.

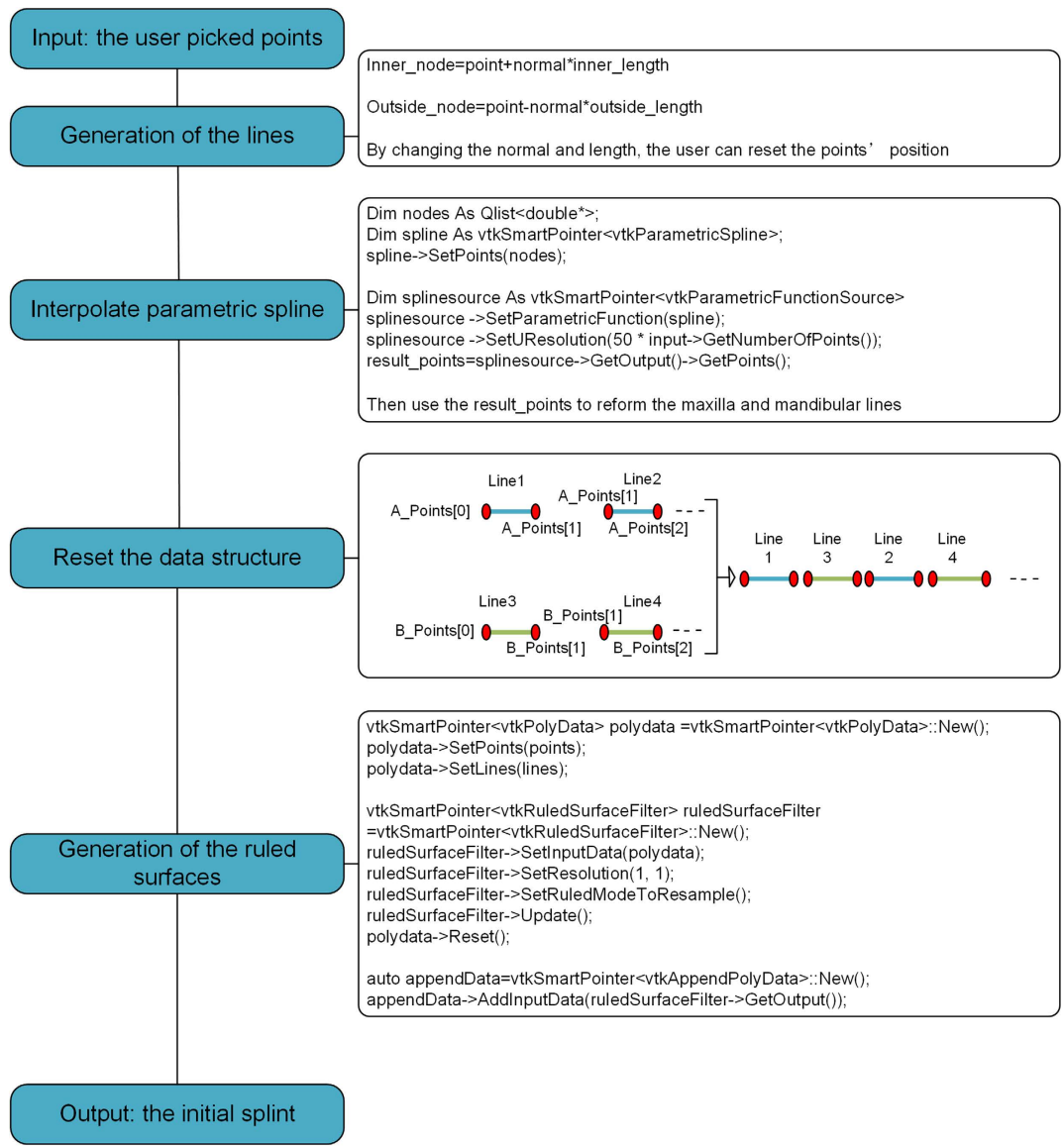

**Figure 3.** The general frame of the generation of initial splint.

4. The module for interference elimination: This module aims at removing the undercut that might exist in the splint. The user can specify the position, direction and distance of the interference elimination. Specified planes are utilized to cut the model on the specific direction. Based on the contour of the cutting, 3D images can be generated under the principle that the upper layer should contain the geometry information of the lower layer. A model then can be reconstructed from the image data by applying the Marching Cube (MC) algorithm[16]. Finally, a Boolean operation between the splint with undercuts and the model is used to eliminate the interference.

The core modules of EasySplint are the generation of the splint base, Boolean Operation and interference elimination.

**The Generation of the Initial Splint.** The design of the initial splint aims to construct a model with customized shape and size. The generation approach is based on the ruled surface reconstruction. Ruled surface reconstruction is one of the most efficient methods to form a surface for two non-collinear lines. The input of our algorithm is the user-chosen point data. There are two main steps in the generation of the initial splint.

The first step is to use the user-chosen point to generate a line with a certain length and direction. The line's position, length and normal can be adjusted manually to ensure the initial splint's shape and size. The lines are shown in Fig. 1.

The second step is to use the lines' nodes to reconstruct the ruled surfaces. Two lists (up list and down list) are used to store the lines which are generated by the points picked from maxilla model and mandibular model. The data structure can be shown as follow:

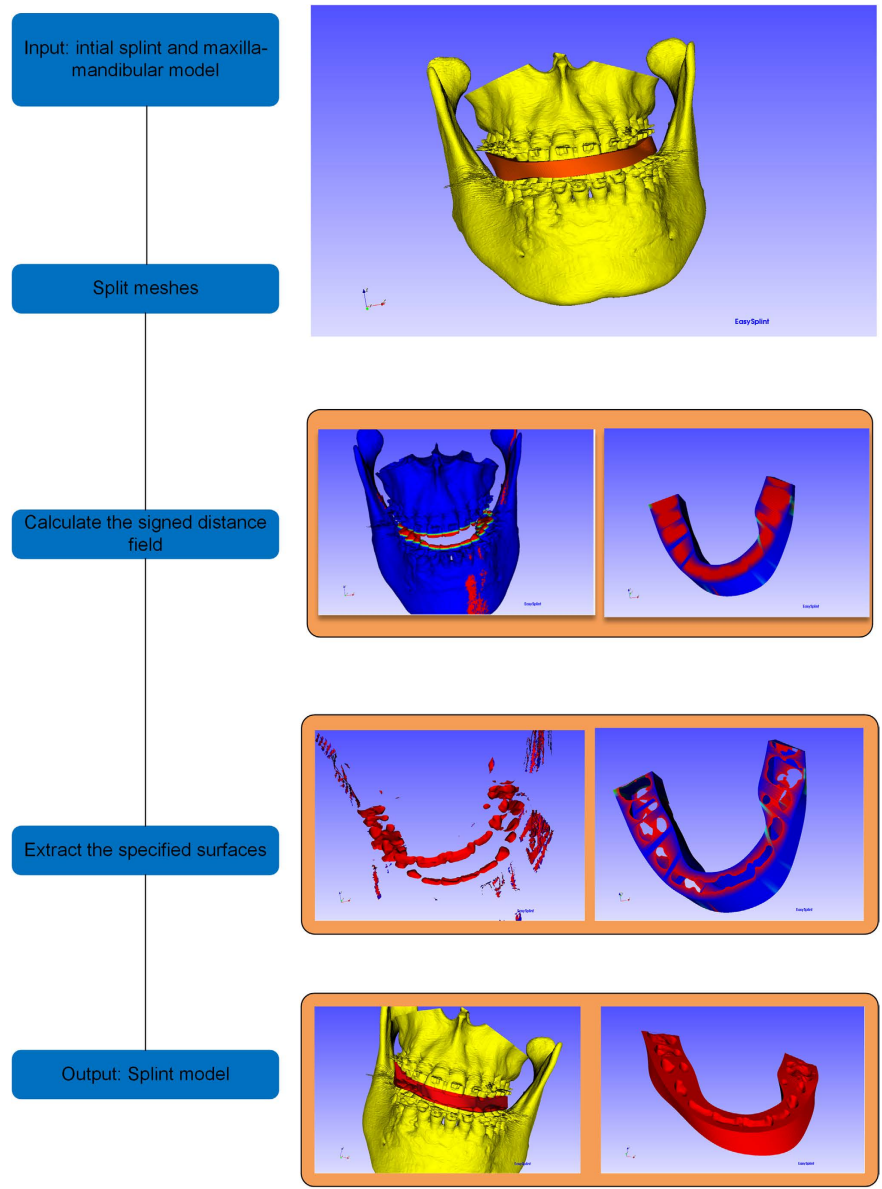

**Figure 4. The general frame of the generation of final splint.**

```
Line Structure
{
    double normal;
    double inner_point[3];
    double outside_point[3];
    double inner_length[3];
    double outside_length[3];
}
```

Every two points from the up list and every two points from the down list can form a ruled surface, which is shown in Fig. 2. By this mean, a number of ruled surfaces are able to be reconstructed and appended as the initial splint. In order to make the shape of the splint smoother, we firstly interpolate the points into parametric splines and then sample enough points to reconstruct the ruled surface. The general frame of the generation of initial splint can be show in Fig. 3.

**The Boolean operation.** A Boolean operation is used to define a third model C by computing model A and model B. There are three common Boolean operations, which are union, intersection and difference.

The Boolean operation used in the EasySplint is difference, between the initial splint and the maxilla-mandibular model. The algorithm of Boolean operation is based on some vtk classes, and can be divided into three main steps, which is shown in Fig. 4 and can be described as follow:

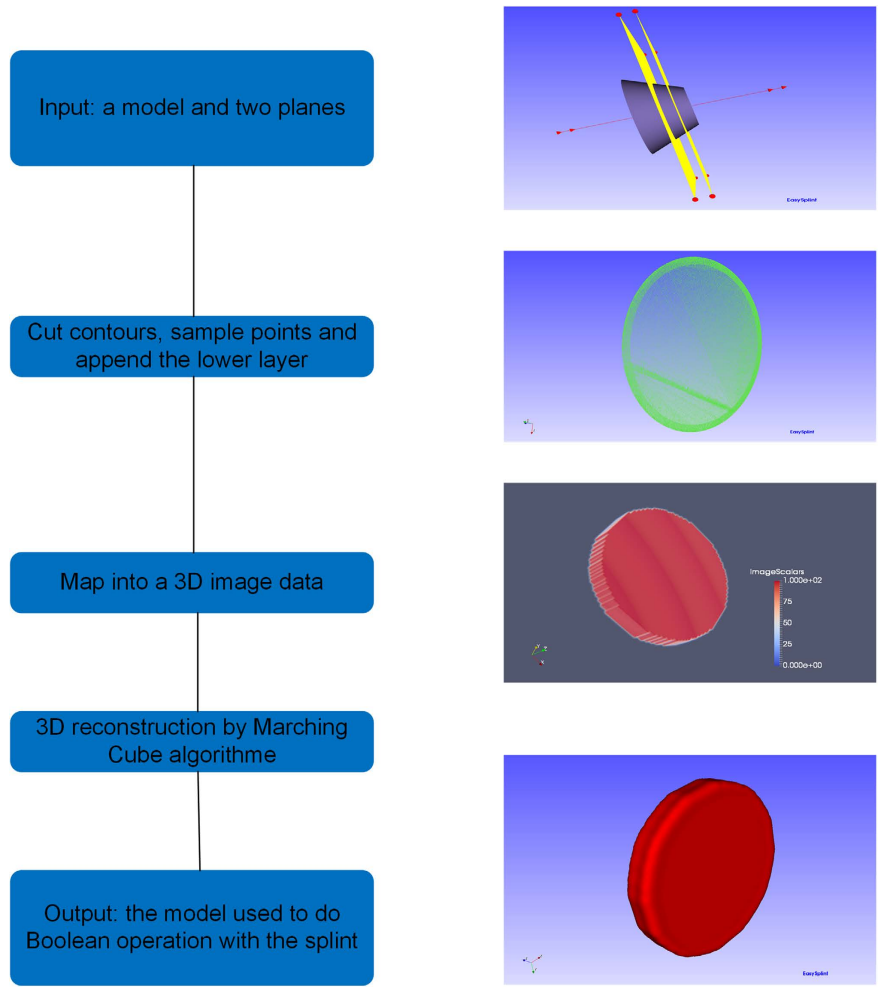

**Figure 5.** The general frame of the interference elimination.

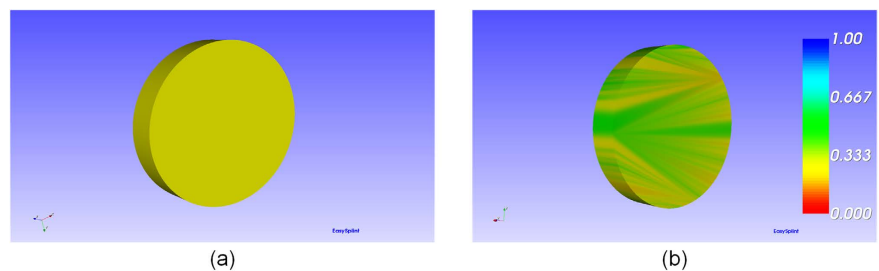

**Figure 6.** The error of the reconstructed model: (**a**) the standard model designed by SolidWorks. (**b**) the error between the standard model and reconstructed model.

1. Split meshes: the main part of Boolean operation algorithm is to identify the intersection surfaces between two surface meshes.
2. Calculate the signed distance field: the distance from points in one surface to another surface is calculated. The distance is defined as the distance between the point of one surface and the nearest point of the other surface. If the point is inside the space bounded by the mesh, then the distance is negative. In other words, if the point is outside the mesh, then the distance is positive.
3. Extract the specified surfaces: with the signed distance field, cells of the surface with certain value of the distance can be extracted. The Boolean operation difference can be defined as the cells of model A whose points are non-negative distance from model B combined with the cells of model B whose points are non-positive distances from A.

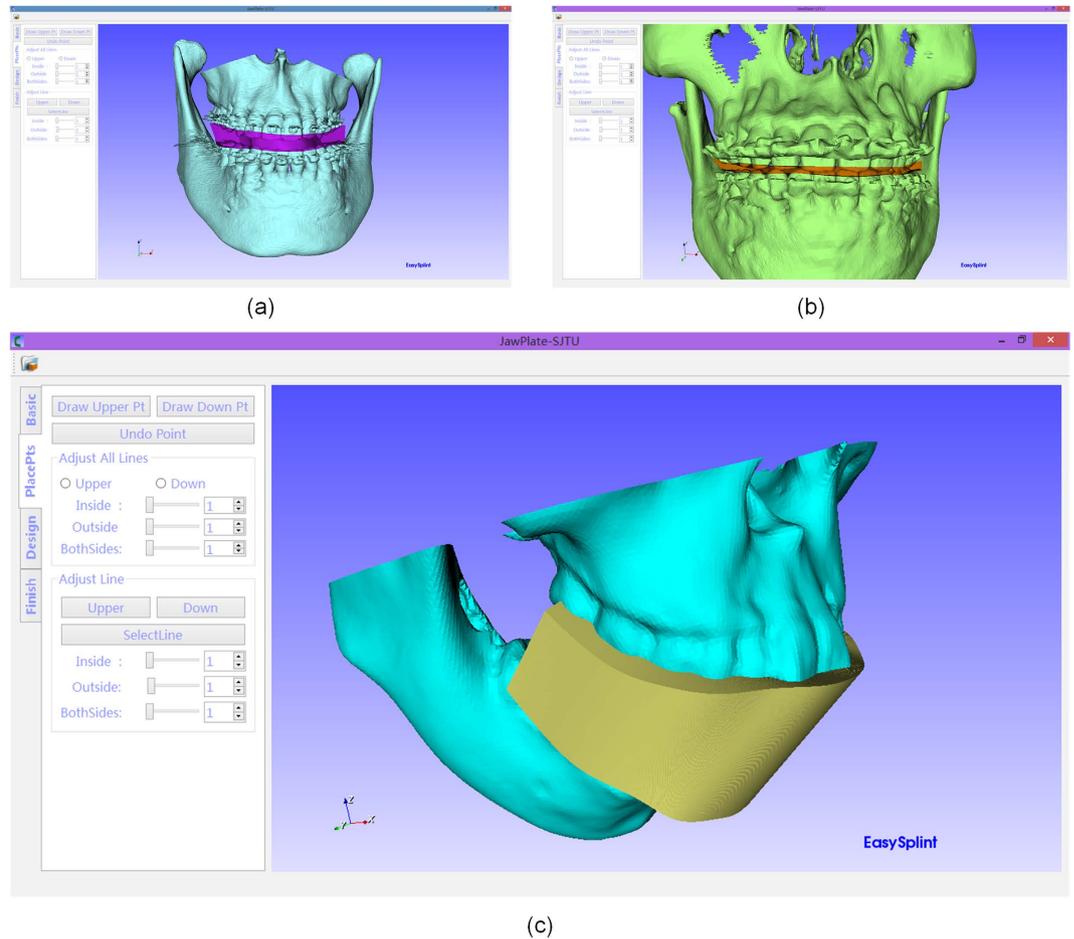

**Figure 7.** The phantom experiments: (**a**,**b**) are the splints the designed splint for the orthognathic surgery and (**c**) is for the orthognathic postoperative opening-mouth training.

**The Interference Elimination.** As the splint is generated based on the shape of the maxilla-mandibular model, undercut may exist in the splint when the teeth of the patient are big-end-up. On that condition, the splint won't fit the patient as the existence of the interference between the teeth and the splint. A novel method is proposed to eliminate the interference. The main idea of the algorithm is based on 3D reconstruction and Boolean operation. A temporary model is reconstructed based on a 3D points volume firstly, which is created according to the space information when the original model moving along the specified direction for interference elimination. Then Boolean operation between the temporary model and the splint is applied to eliminate the part of the splint which causes the interference. As the Boolean operation has already been discussed above, this section will mainly focus on how to reconstruct the temporary model.

Marching Cubes algorithm (MC algorithm) is adopted to accomplish the reconstruction of the model. The main steps can be described as the generation of the contour, the construction of the image data, the image data processing and the surface reconstruction. The typical structure which will cause interference is cone. A cone with big-end-up is used as the original model to illustrate the method works, which is shown in Fig. 5. The standard model in Fig. 6(a) is a surface model designed using the software of SolidWorks with the exact radius of the first contour cut by the plane. The comparison between the standard model and the reconstructed one was evaluated by computing the distance errors of these two models, and the result of color map is shown in Fig. 6(b). The mean distance error is calculated to be 0.365 mm, which is precise enough in clinical application[14]. The details of the method for measurements are described in the second step of the subsection of "Boolean operation".

The first step is to generate the contour of the model. The contour contains the information of the space extend of the model moving along the direction of the undercut removal. A plane is used to cut the model to get the contour, and then enough points are sampled inside the contour. By offsetting the cutting plane, an array of contours of the model can be acquired. Based on the principle that the upper layer should contain the geometry information of the lower layer, we append the upper layer contour and lower lay contour to form a volume of points.

The second step is to map the point volume into a 3D image data. The size of the point volume will be calculated firstly in order to define a 3D image data. Based on the points from the volume, the correspondent position of the 3D image data will be assigned a non-zero scalar value, whereas other positions' scale value will be zero.

Image dilation and image erosion are applied in the xyz direction to improve the quality of the 3D image data. Then the temporary model will be reconstructed with the MC algorithm based on the 3D image data.

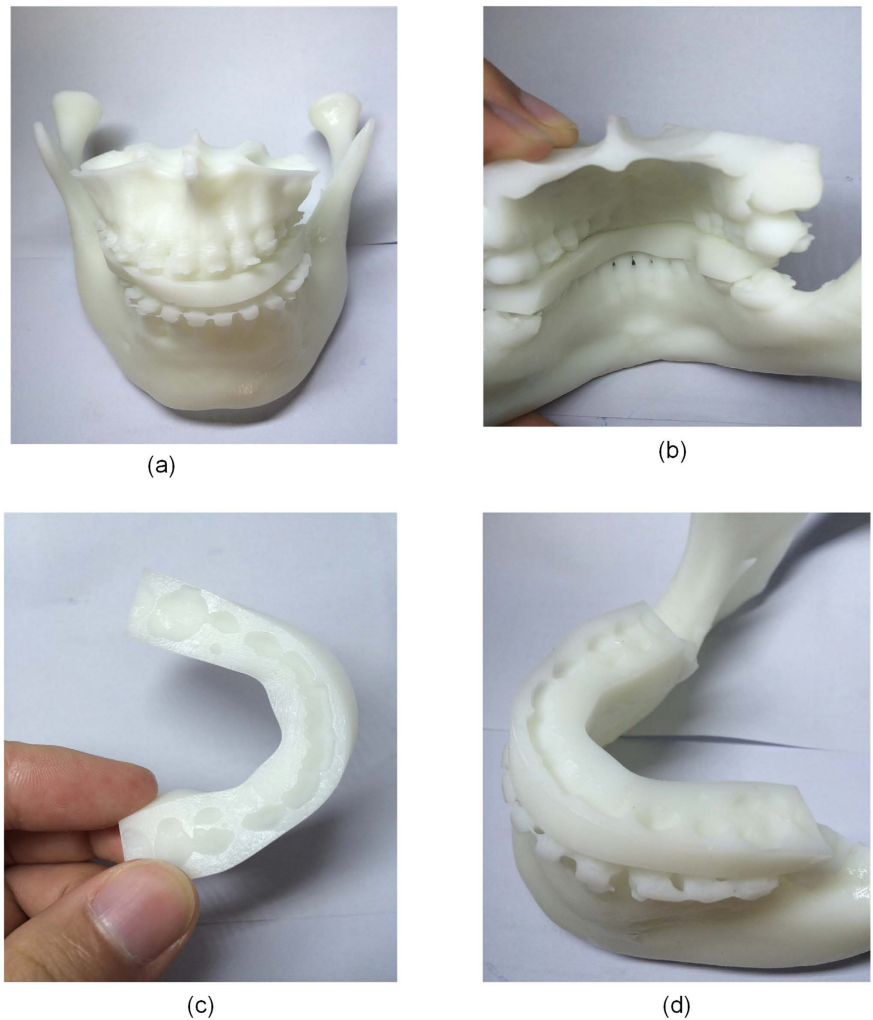

**Figure 8.** The phantom experiment: (**a**,**b**,**c**,**d**) are the different views of the fabricated splint of one case (via Eden260 VS, Stratasys Ltd).

| Index | Time for the generation of the initial splint(s) | Cells of the initial splint | Points of the initial splint | Cells of the maxilla-mandibular model | Points of the maxilla-mandibular model | Time for Boolean Operation(s) | Whole time(min) |
|---|---|---|---|---|---|---|---|
| 1 | 0.132 | 2640 | 1398 | 543530 | 271337 | 55.325 | 12 |
| 2 | 0.148 | 2560 | 1358 | 324246 | 161832 | 33.877 | 11 |
| 3 | 0.061 | 2560 | 1358 | 86866 | 43433 | 7.575 | 10 |

**Table 1. The analysis of three phantom experiments.**

## Results

A general framework of the splint design was introduced and several algorithms including the ruled surface generation, Boolean operation and interference elimination were presented. On the basis of these algorithms, a software named EasySplint was developed under the platform of Microsoft Visual Studio 2010 (Microsoft, Washington, USA). Some famous open source toolkits including VTK (Visualization Toolkit, an open-source, freely available software system for 3D computer graphics, image processing and visualization, http://www.vtk.org/) and Qt (a cross-platform application and UI framework, http://qt-project.org/) were involved. Several cases of customized splint design for the orthognathic surgery were conducted using EasySplint. With the manually drawn points indicating the lines with a specified direction and length, the splint can be generated automatically and rapidly. If the maxilla-mandibular model has some big-end-up parts, the interference elimination can be conducted.

Three phantom experiments were conducted, of which two were for the orthognathic surgery and one is for the orthognathic postoperative opening-mouth training shown in Fig. 7. The fabricated splint of one case shown in Fig. 8 demonstrated the designed splint can fit the maxilla-mandibular model well. All the phantom experiments were conducted on a PC with Intel Core i7-4710MQ with a 2.50 GHz CPU, 8 GB memory and a 64-bit

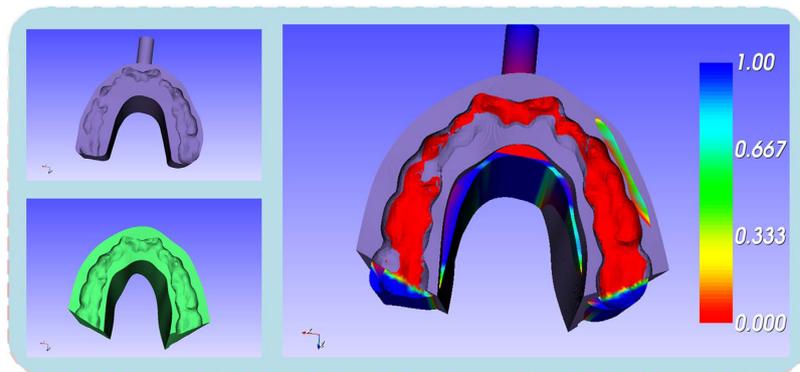

**Figure 9. The comparison of dental splints designed using the ProPlan CMF 1.3 (top left corner) and the EasySplint (lower left corner).**

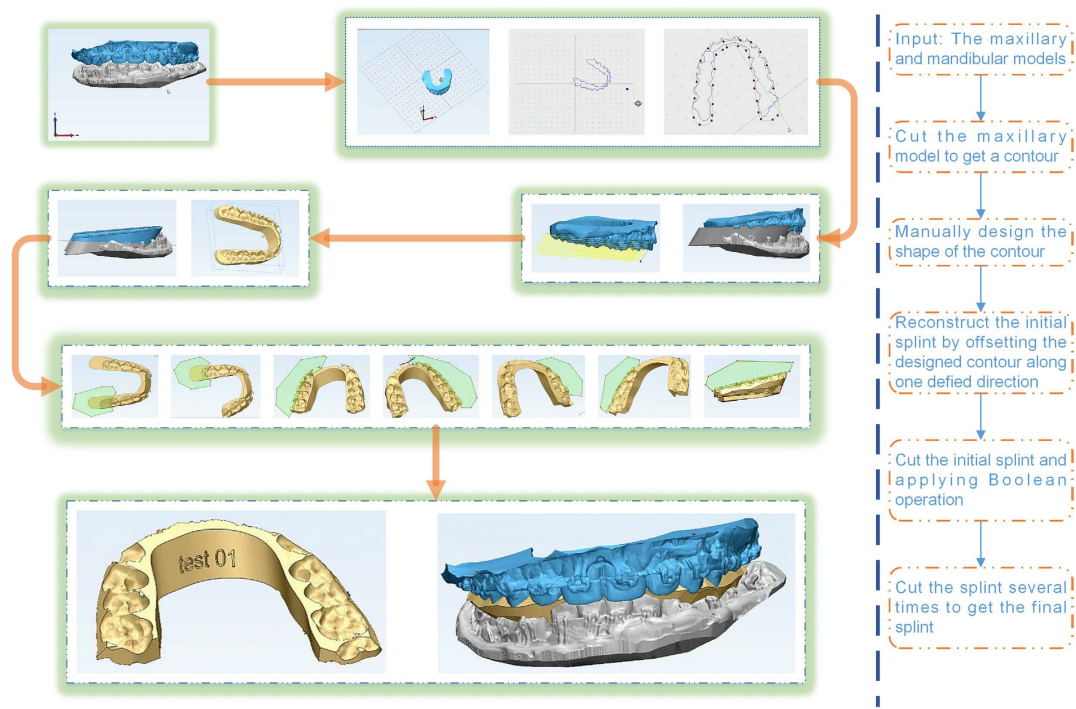

**Figure 10. The workflow of dental splint design using 3-matic.**

Windows 8 operating system. Table 1 shows the property of the input maxilla and mandibular model and the corresponding computing time of each step during the design of the splint. In this table, time of the generation of the splint is the sum of 'Lines adjustment', 'Generation of the initial splint', and 'The Boolean operation'. For all examples, the overall time for the design of the splint is about 10 minutes.

In terms of accuracy, we made comparison study on the splint design between our method and commercial softwares. Previously, clinical experiments of the dental splints designed with commercial softwares, such as ProPlan CMF, Unigraphics NX, VisCAM, etc., were conducted and reported by our research group[17]. It shows that the difference between the planned and the actual bony surgical movement at the edge of the upper central incisor was $0.50 \pm 0.22$ mm in sagittal, $0.57 \pm 0.35$ mm in vertical, and $0.38 \pm 0.35$ mm in horizontal direction (midlines) for 15 patients. Then, on the basis of the same data, the dental splints were designed using EasySplint, and the distance errors between these two kinds of splints were computed and the color map is shown in Fig. 9. The results show that there was no difference between them in the occlusal area of the dentition, which means that the accuracy of the splint designed using EasySplint can be guaranteed as same as that of using commercial softwares. Nevertheless, it will be further validated with clinical experiments in the future study.

### Discussion

The commercial softwares are not specially developed for dental splint design, so the procedures for dental splint design tend to be complicated. Take 3-matic (Materialise®, Leuven, Belgium), one of the most commonly adopted softwares for template design, as an example: the time-consuming workflow of dental splint design is shown in Fig. 10 and it usually takes 0.5~1 hour to accomplish the whole design procedure. Other commercial softwares, such as Imageware, UG, Pro/E, ProPlan CMF etc. are even more complicated, and the data import and export among them are inevitable due to that the functions of each software are limited and the whole design can only be accomplished through the combination of those softwares. It also means it requires high level of the engineering background for the user to grasp all of those softwares and get very familiar with the interface among them. The rate limiting steps for the design time in those softwares are complicated manual operations of geometry design including combining surfaces, repairing and de-featuring, remeshing, modifying and editing, etc.

The EasySplint software can provide an easy and efficient approach to design a customized splint for orthognathic surgery based on the 3D maxilla-mandibular model of the patient. The shape and position of the initial splint is planned by user's specified points. By applying the Boolean operation between the initial splint and the maxilla-mandibular model, the final splint is produced automatically. The interference elimination has been taken consideration based on a novel method using the Marching Cubes algorithm.

The conducted phantom experiments indicate that the splints designed by the EasySplint software are suitable for the orthognathic surgery or the open-mouth exercising. As shown in the Table 1, the time of the splint design is about 10 minutes whereas it need 30~60 minutes using the traditional softwares.

Currently, the Easy Splint software is just in a preliminary stage and some functions will be improved in the future work. Firstly, if there are artifacts in the original CT images, the 3D-reconstructed maxilla- mandibular model will not be accurate enough for the splint design. Under this condition, we will use laser scanning technology to obtain the precise dentition data of the cast model, and then register the CT data and the laser-scanned data using the method we described in the ref. 18. Secondly, the sharpness of edges of the splint hasn't been taken consideration in EasySplint. Since the splint with shape edges may injure the soft tissue of the patient in orthognathic surgery, or cause uncomfortable feeling for the patient in postoperative opening-mouth training, how to fillet the splint still need further study. Thirdly, if the size of the original model for interference elimination is too large, the operating system will not be able to allocate enough memory. In this case, we will develop a novel 3D reconstruction and mesh decimation algorithms to decrease the error and improve the quality of the temporary model for interference elimination. Furthermore, clinical applications will be conducted in the near future to demonstrate its feasibility and reliability.

## Acknowledgements

This work was supported by Natural Science Foundation of China (Grant No.: 81511130089), Foundation of Science and Technology Commission of Shanghai Municipality (Grant No.: 14441901002, 15510722200, and 16441908400) and SJTU - KU Leuven Fund.

### Author Contributions
X.C. and X.L. conceived of the project, and designed the framework of the software. X.C., X.L., and J.E. proposed the new algorithms, developed the software and performed the experiments. X.C., X.L., L.X. Y.S., C.P., and J.E. wrote the paper. All authors discussed the results and commented on the manuscript.

### Additional Information
**Competing financial interests:** The authors declare no competing financial interests.

**How to cite this article**: Chen, X. *et al.* Development of a computer-aided design software for dental splint in orthognathic surgery. *Sci. Rep.* **6**, 38867; doi: 10.1038/srep38867 (2016).

**Publisher's note:** Springer Nature remains neutral with regard to jurisdictional claims in published maps and institutional affiliations.